\documentclass[twocolumn,floatfix]{revtex4-1}

\usepackage{amsthm}
\usepackage{amsfonts}
\usepackage[english]{babel}
\usepackage[utf8]{inputenc}
\usepackage{mathtools}
\usepackage{physics}
\usepackage{float}
\usepackage{xcolor}
\usepackage{graphicx}
\usepackage[left=23mm,right=13mm,top=35mm,columnsep=15pt]{geometry} \usepackage{adjustbox}
\usepackage{placeins}
\usepackage[T1]{fontenc}
\usepackage{lipsum}
\usepackage{csquotes}
\usepackage{hhline}
\usepackage{multirow}
\usepackage{enumerate}
\usepackage[pdftex, pdftitle={Article}, pdfauthor={Author}]{hyperref}
\usepackage{cleveref}
\usepackage[bottom]{footmisc}

\usepackage{makecell,tabularx}
\setcellgapes{3pt}

\usepackage{caption}
\usepackage{subcaption}
\usepackage{listings}
\usepackage{array}
\usepackage{arydshln}
\newcolumntype{P}[1]{>{\centering\arraybackslash}p{#1}}
\newcolumntype{M}[1]{>{\centering\arraybackslash}m{#1}}

\begin{document}

\title{Networks of triblock ellipses: from kagome order to disordered porous networks}

\author{S.~Wagner$^1$, G.~Kahl$^1$ and C.~Karner$^1$}
\email[Correspondence email address: ]{carina.karner@tuwien.ac.at}
\affiliation{$^1$Institute for Theoretical Physics, TU Wien, Wiedner Hauptstra\ss e 8-10/136, A-1040 Wien, Austria}

\date{\today}

\begin{abstract}
In colloids, anisotropy in either shape or particle interactions can steer assemblies away from close-packed monolayers and towards more intricate structures. In this study we utilize both shape and interaction anisotropy to stabilize porous surface assemblies of triblock elliptical particles, whose surfaces are decorated with two attractive patches at opposite poles. The resulting structures display a tunable degree of crystalline kagome order, controlled by quenching the temperature and/or changing the chemical potential. In particular, we find evidence that nucleation into the kagome lattice proceeds via a one-step pathway, in contrast to the two-step route widely reported in spherical triblock systems, making triblock ellipses an interesting model system for studying how self-assembly pathways change with particle shape. Beyond ordered crystalline structures, we find disordered porous networks with complex internal geometry. To characterize this range of assemblies, we introduce a particle-patch graph that explicitly captures the localization of bonds at distinct patches. Bonding motifs follow directly from this representation, while the loops, that we identify as the pore boundaries, can be determined from the graph's planar embedding. We detect all loops as the face boundaries in the planar embedding by means of a face-walking algorithm, distinct from shortest-path approaches used in ring detection. Together with a new local and global order parameter measuring the loop network's regularity with respect to a selected loop size, this framework connects microscopic bonding motifs to mesoscopic pore properties and quantifies the tunable order of these porous assemblies.
\begin{description}
\item[Usage]
Electronic Supplementary Information (ESI) available. 
\end{description}
\end{abstract}

\maketitle

\section{Introduction}\label{sec:intro}
Colloidal particles can be functionalized with attractive regions at selected positions on their surface, creating patchy particles with directional and valence-limited interactions \cite{Solomon2007, BianchiBlaak2010, LiZhouHan2016, BianchiIPC2017, ravaine2017synthesis, zhang2017janus}. By controlling the number, position and extent of these patches, local bonding geometries and assembled structures can be programmed \cite{millan2014self, Knappe2021, Karner2019, GranickChen2011, he2020colloidal, Liu2024,Tang2023, li2018general,ShahZhang2015,shah2013synthesis, zhang2011synthesis, Liu2012}.
In particular, patch localization enables assembly into open and porous networks that are otherwise difficult to stabilize with purely isotropic attractions: while uniformly attractive particles minimize their energy by maximizing their coordination, patchy particles can saturate their attractive interactions already in low-coordinated structures. This provides access to open crystals and networks whose topology and pore geometry can give rise to distinct optical, mechanical, transport and wetting properties \cite{Lam2026,Bushev2022,Waterhouse2008,Dyogtyev2010,Lee2012,Lubensky2009}. 

A prominent example is the self-assembly of triblock patchy particles into the two-dimensional kagome lattice. Spherical particles with two oppositely positioned attractive caps, sufficiently large to accommodate two bonded neighbors each, generate an open lattice \cite{GranickChen2011,Romano2011}.
The resulting kagome geometry is specifically interesting for its gap in the photonic band structure and for its low-coordinated, nearly isostatic mechanical structure \cite{Dyogtyev2010,Lee2012,Lubensky2009}. Its assembly pathway is equally distinctive: for spherical triblock particles, formation of the kagome lattice has been reported to proceed through a two-step process in which particles first form chains and disordered networks that subsequently rearrange into the open lattice \cite{GranickChen2011,Bahri2022,Gharibi2024,Romano2011}.

In this contribution, we study an aspheric extension of this triblock system: by elongating the particle along one axis we generate triblock ellipses with aspect ratio $\kappa=2$ and two attractive patches placed on the ellipses' poles, each capable of accommodating up to two bonds. This raises a natural question: how does shape anisotropy affect both the assembled structures and the pathways through which they form?
We find that, as for their spherical counterparts \cite{Romano2011, Bahri2022, GranickChen2011}, the introduction of patches pushes ellipses away from close-packing \cite{CuestaFrenkel1990, BautistaOdriozola2014, Madivala2009, Foulaadvand2013, WagnerBaumketner2024} towards porous structures. 

We study the patchy ellipses self-assembly across a range of temperatures and chemical potentials and find that the kagome lattice assembles within a specific region of this parameter space. This is -- to the best of our knowledge -- the first observation of a kagome lattice assembled from patchy ellipses. We further find evidence for a one-step nucleation pathway from the dilute phase, without the distinct chain intermediate reported for spherical triblock particles \cite{Romano2011, Bahri2022, GranickChen2011}.

So far, computational as well as experimental work on patchy ellipses \cite{li2018general,ShahZhang2015,shah2013synthesis, zhang2011synthesis, Liu2012} has focused on the assembly of clusters and finite structures. For elliptic particles in two and three dimensions with Janus \cite{li2018general, shah2013synthesis, Liu2012, ShahZhang2015}, one-patch \cite{li2018general, shah2013synthesis} and two-patch \cite{li2018general, zhang2011synthesis} architectures,  micelles, vesicles, chains and fibres have been stabilized. From this perspective, our work broadens the existing body of literature to include the study of the assembly of extended structures across a whole range of thermodynamic conditions. 
In our system, deeper quenches to lower temperature and/or higher chemical potential lead to increasingly disordered particle networks. As the disorder increases, we find that the kagome lattice, appearing as a regular network with pores bounded by loops of six particles, develops defect structures that appear different from the irregular pore networks of deeply quenched spherical triblocks studied so far \cite{eslami2018solid, schubert2025self, GranickChen2011, Romano2011, Bahri2022}. In our system, defect structures are more reminiscent of those observed in trivalent patchy particles \cite{SwinkelsNoya2023, SwinkelsSchall2023}: triblock ellipses as well as trivalent spheres exhibit characteristic defects of five- and seven-membered loops, within the otherwise six-membered loop networks.

Moving down in length scale, similar defects also occur in atomic and molecular networks, where such loops are commonly referred to as rings. Vitreous silica bilayers exhibit rings of various sizes \cite{Lichtenstein2012, Buchner2017, Gao2016, AltmanMalashevich2016, kumar2014ring, zhang2018phase, ebrahem2020vitreous}, while five- and seven-membered rings are prominent features of disordered two-dimensional water networks \cite{koga2000first, nie2010pentagons,yuan2025atomic} and amorphous carbon monolayers \cite{toh2020synthesis, eder2014journey,van2012insulating}. These structural similarities reflect a shared organizing principle: directional bonding and limited valence constrain local connectivity, with covalent and hydrogen bonds playing a role analogous to that of attractive patches in colloidal systems.

Changes in network order can substantially affect material properties: increasing disorder reduces the stiffness and tensile strength of two-dimensional silica \cite{zhang2018phase, ebrahem2020vitreous}, while strong topological disorder can induce insulating behavior in graphene \cite{van2012insulating}. These structure–property relationships motivate a quantitative characterization of network topology. Here, graph-based ring statistics provide an established approach to characterizing ordered and disordered molecular networks \cite{Cormack2002,Goetzke1991}, offering a natural starting point for us to analyze the analogous structures in colloidal systems.

However, for patchy ellipses a simple particle graph, abstracting particles as nodes and bonds as edges, is insufficient: because each patch can participate in multiple bonds, a single node per particle cannot distinguish bonds formed at the same patch from those distributed across opposite patches. We therefore introduce the particle--patch graph, in which both particle cores and individual patches are represented explicitly as nodes. This novel representation allows us to identify local bonding motifs, branching structures, and loops within a unified framework.

Based on the detected loops, we introduce local and global loop-order parameters that quantify the prevalence and regularity of selected loop sizes. Together with the identification of local bonding motifs, these measures allow us to follow both the emergence of kagome order during assembly and its variation across thermodynamic conditions. We use this framework to characterize the progression from ordered kagome lattices to heterogeneous porous networks and to resolve the microscopic pathway by which the lattice forms from the dilute phase. The particle--patch graph therefore provides the optimal framework to study the two central questions of this work: how the particle's deviation from circular shape affects the emerging structures and the pathways through which they emerge.

The remainder of this paper is organized as follows: In Section~\ref{sec:methodsandmodels}, we introduce the patchy ellipse model and the simulation methods used to study its self-assembly. Section~\ref{sec:analysis} introduces the particle-patch graph and the methods defined on it to identify bonding motifs, loops, and loop order. In Section~\ref{sec:results}, we present our results, focusing on two cross-sections of the parameter space, one at constant temperature and one at constant chemical potential, and characterize the emerging structures in terms of their loop order. We then turn to very low chemical potentials, where we find that self-assembly proceeds via nucleation. In section~\ref{sec:conclusio} we summarize our findings and discuss open questions.

\section{Model and Simulation Methods} \label{sec:methodsandmodels}
\subsection{Patchy Ellipse Model}\label{sec:pm}
We study a patchy ellipse model combining two types of anisotropic interactions in a two dimensional setup: patches and elliptic shape \cite{Wagner2019}. The hard ellipses with aspect ratio $\kappa=2$ interact via a hard-body interaction. Here, the aspect ratio is defined as $\kappa=\frac{a}{b}$ with $a$, $b$ denoting the major- and minor-axes respectively, and the unit of length $\sigma=2\sqrt{ab}$. For hard ellipses the mutual interaction of two particles can easily be identified via the Vieillard-Baron overlap criterion \cite{VieillardBaron1972,VieillardBaronThesis}. The two patches, placed at the ellipse's poles, i.e. the two endpoints of its major axis, are of Kern-Frenkel type \cite{Kern2003}, appearing as attractive disk segments with half-opening angles $\theta=45^{\circ}$ and interaction range $\delta=0.17\sigma$. With consideration of the interaction range $\delta$, the patch half-opening angle $\theta=45^{\circ}$ allows up to two bonds on each patch. We use the depth of the square-well potential as the unit of energy, i.e. $\epsilon=1$.

\subsection{Simulation Methods}\label{sec:sm}
Surface assembly of patchy ellipses is simulated using Monte Carlo (MC) simulations in the grand-canonical ensemble, specified by temperature $T$, volume (i.e, area) $V$ and chemical potential $\mu$. Particle insertion and deletion moves allow the surface coverage to vary with the imposed chemical potential $\mu$. To enable collective rearrangements of bonded particles, we employ cluster translation and rotation moves using the static linking scheme of Ref.~\cite{Whitelam2007}.\medskip

For control parameters in the simulations we use dimensionless thermodynamic quantities, such as the dimensionless temperature $T^*=k_{\textrm{B}}T/\epsilon$, dimensionless chemical potential $\mu^{\ast} = \mu/(k_{\textrm{B}}T)$, and dimensionless density $\rho^{\ast} = \rho/\rho_{\textrm{cpd}}$, where $\rho_{\textrm{cpd}}$ is the close-packed density of hard ellipses, $\rho_{\text{cpd}}=\frac{2}{\sqrt{3}}\sigma^{-2}$. We report all numerical values using these dimensionless forms but for notational clarity write $T$, $\mu$, and $\rho$ in place of $T^{\ast}$, $\mu^{\ast}$, and $\rho^{\ast}$, respectively.
\medskip

We investigate a grid of 104 thermodynamic state points spanned by temperature and chemical potential. The studied state points extend over eight equidistantly spaced temperatures $T\in[0.055,\ldots,0.16]$, with $\Delta T=0.015$, and thirteen equidistantly spaced chemical potentials $\mu\in[-7.0,\ldots,5.0]$, with $\Delta\mu=1$. At each state point we perform twelve independent simulations resulting in a total of 1248 trajectories.\medskip

Each simulation begins from a dilute particle gas with density $\rho^{\ast}\approx0.004$, which is instantaneously quenched to the target thermodynamic conditions defined by $(\mu,T)$. A cycle in our MC simulations is defined as $N^{\rm cp}$ proposed MC moves, with $N^{\rm cp}=1296$ corresponding to the number of particles in the simulation box at close packing. The simulation protocol starts with $2\times10^{6}$ cycles during which the maximum displacements of the translational and rotational MC moves are adjusted to obtain acceptance ratios between $0.2$ and $0.5$. Since the maximum displacements of moves are modified during this stage, these startup-cycles are discarded from the subsequent analysis.\medskip

In the subsequent production run, the maximum displacements are kept fixed at the values obtained during tuning. The configuration is recorded every $2\times10^{3}$ cycles. A single-particle translation or rotation move is proposed with probabilities $p_{\rm t}=p_{\rm r}=0.38$, cluster translations and rotations with probabilities $p_{\rm tc}=p_{\rm rc}=0.095$ and particle deletions and insertions with probabilities $p_d=p_i=0.025$. Depending on the thermodynamic state point, the total number of sampled configurations after the discarded cycles range from roughly $n_{\text{conf}}=1\times10^{6}$ to $2\times10^{7}$ cycles, i.e. sample sizes of $5\times10^2$ to $1\times10^4$ configurations.
For the time-based nucleation analysis we refer to one cycle as mc-time.

\subsection{Analysis of simulation trajectories}\label{sub:plateau}
To determine which simulation runs have reached a final assembly state, we analyze the total energy $E_{\mathrm{tot}}(t_i)$ along the trajectory, with $i=1,\dots,n_{\text{conf}}$ indexing the configurations. We test whether the trajectory reaches a plateau or instead exhibits a slow, monotonic drift, which we refer to as creeping. Since the onset of a plateau appears as a change in slope on short time scales, whereas creeping only becomes apparent over long time intervals, the test consists of two steps.

In the first step, we determine the onset $t'_0$ of the plateau. We smooth the trajectory with a moving average acting as a low-pass filter, with a width between $50\tau_{\mathrm{mc}}$ and $500\tau_{\mathrm{mc}}$ ($\tau_{\mathrm{mc}}=2\times10^3$ cycles) that covers less than a quarter of the trajectory. We then determine the slope $k$ of the smoothed trajectory by linear regression over a sequence of overlapping time windows of half the filter width. Proceeding backwards in time, $t'_0$ is the start of the last window that meets $|k|<k'_{\text{tol}}$, with $k'_{\text{tol}}=0.016\,k_{\mathrm{B}}T/\tau_{\mathrm{mc}}$.

Since short time windows cannot reveal creeping, in the second step we apply a linear regression to the unfiltered tail $[t'_0,t_{n_\text{conf}}]$ and classify the trajectory as creeping, i.e., not plateauing, if both $|k|>k''_{\text{tol}}$, with the stricter tolerance $k''_{\text{tol}}=0.01\,k_{\mathrm{B}}T/\tau_{\mathrm{mc}}$, and $R^2\geq 0.1$. The low $R^2$ threshold accepts the fluctuations of the raw data, while still requiring the linear trend to be statistically noticeable. For trajectories without creeping, the plateau is reached at $t_0=t'_0$. Trajectories for which the first step finds no plateau onset before the last 2\% of the trajectory are instead tested over their last 20\%, and assigned $t_0$ at the start of this interval if no creeping is detected. Details are given in the Electronic Supplementary Information (ESI).

\section{Analysis Methods}\label{sec:analysis}
Here we present analysis methods that rely on information about the particle bonds encoded within a graph. With this graph-based approach we analyze the patchy ellipses network structure across scales, spanning from local bonding motifs up to the open network structures characterized mainly by loops. 
First, we introduce a graph representation, suitable for general patchy particle networks, then proceed to the detection of bonding motifs and loops, and finally introduce loop order parameters as measures of local and global order with respect to a selected loop size. We refer to the Appendix for a relevant introduction to graph concepts and to Ref.~\cite{graphtheory} for a more comprehensive treatment thereof.

All analysis methods presented here are made publicly available via the python-package \texttt{patchynet} \cite{patchynet}, a software package based on \texttt{networkx} \cite{networkx} and specifically developed for the graph-based structural analysis of patchy particle networks in two dimensions. Although developed for networks of patchy ellipses, the methods are purely graph-based and therefore universally applicable to any two dimensional network built from monomeric units with fixed valency.

\subsection{Graph Representations}
\begin{figure}
    \includegraphics[width=.8\linewidth]{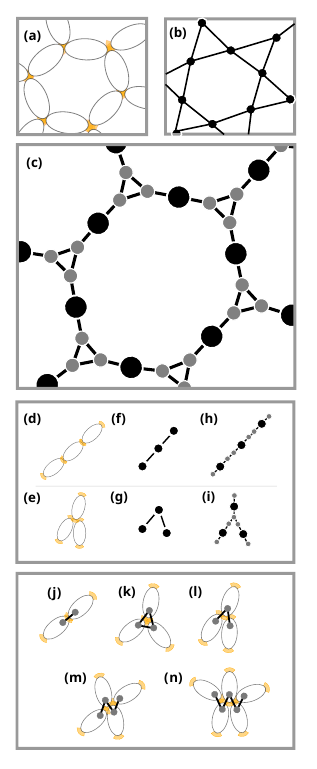}
    \caption{(a) A network of patchy ellipses (elliptic cores in gray, patches in orange) abstracted by (b) the particle graph as well as (c) the particle-patch graph, with a particle represented as a core node (black) connected to its two patch nodes (gray), and a bond represented as an edge connecting two patches; A particle with two neighboring particles connected (d) by one bond per opposing patch and (e) by two bonds at the same patch; (f-g) the associated particle graphs and (h-i) the associated particle-patch graphs; (j-n) typical bonding motifs and their corresponding patch graphs: (j) pair, (k) triangle, (l) open triangle, (m) N-motif and (n) W-motif.}
    \label{fig:graph}
\end{figure}

Conventional methods in particle network analysis use a graph representation that we call the particle graph. This graph represents the particle network (see Fig.~\ref{fig:graph}(a)), by a graph in which each particle is represented by a node and a bond between two particles is represented by an edge (see Fig.~\ref{fig:graph}(b)). We find that this straight-forward graph representation is not a suitable choice in the studied system of patchy ellipses. Shown in selected panels of Fig.~\ref{fig:graph}, we illustrate a particle with two bonded neighbors: in one configuration the two neighbors bond to opposite patches (panel (d)), in the other configuration they bond to the same patch (panel (e)). The particle graph, however, maps these two distinct bond configurations onto the same graph (see panels f and g). 

To distinguish these two configurations in the graph representation, we introduce the particle-patch graph, shown in Fig.~\ref{fig:graph}(c), with two types of nodes: a particle core node, that is rigidly connected to its patch nodes. Bonds between patches are represented as edges between the patch nodes. As shown in  Fig.~\ref{fig:graph} (h) and (i) the particle-patch graph now achieves the distinction between the two bond configurations (d) and (e).

\subsection{Bonding Motif Detection}\label{sub:bondingmotifs}
As our particles allow for more than one bond per patch, multiple bonding configurations can appear at a single patch site. We call these configurations the bonding motifs, and want to distinguish between different types. We find that the detection of such bonding motifs follows naturally from the particle-patch graph, when removing all particle nodes from it. What remains are the patch nodes connected by edges, indicating their bonds. In a such defined subgraph, all bonding motifs appear as distinct connected components. With this detection scheme, all bonding motifs appearing in an assembly are directly recovered and classified by the graph structure of the connected component.

Fig.~\ref{fig:graph}(j)-(n) shows the most prevalent of these connected components as they appear within the assembly and therefore also in the subgraph: the pair motif, shown in (j), the triangle motif formed by three patches and three bonds, in (k), the open triangle motif, including again three patches but only two bonds, in (l), the N-motif, in (m) and the W-motif, in (n).

\subsection{Cyclic and Acyclic Subgraphs}\label{sub:subgraphs}
The bonding motifs, introduced in Sec.~\ref{sub:bondingmotifs}, act as the basic building blocks to mesoscale structures such as linear chains, branched chains or loops.

For the detection of loops, we introduce the cyclic subgraph, a subgraph of the particle-patch graph that contains only cyclic components corresponding to loops in the particle network. As can be seen from Fig.~\ref{fig:graph}(c), the particle-patch graph includes two different types of cyclic components: firstly, cycles that correspond to loops and secondly, cycles that correspond to the triangle bonding motif. We are interested in the smallest cyclic subgraph that includes all cyclic components corresponding to loops, as well as triangles connecting these loops. In the picture of the particle network, this means to strip away all particles in linear or branched chains, and thus reducing the network to contain only particles that are part of loops. For the detailed definition of the cyclic subgraph we refer to the Appendix.

\begin{figure}
    \centering
    \includegraphics{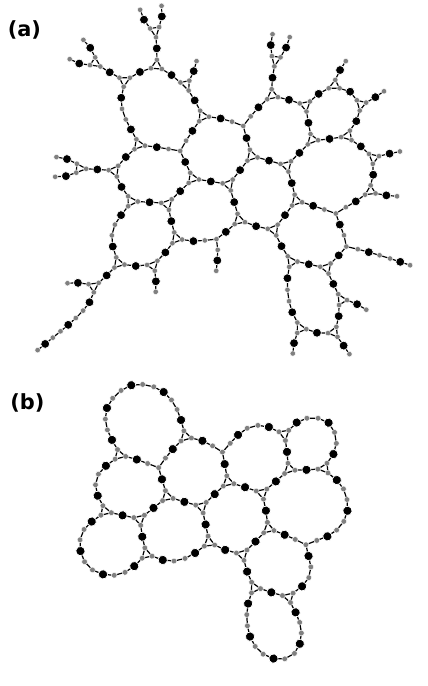}
    \caption{(a) The particle-patch graph and (b) the corresponding cyclic subgraph that represent a network of patchy ellipses forming a nucleus at the early stages of nucleation.}
    \label{fig:subgraphs}
\end{figure}

We find the cyclic subgraph by applying a pruning scheme. In this scheme, we iteratively remove nodes from the graph that do not fulfill the definition of the cyclic subgraph (as defined in the Appendix), until convergence to the cyclic subgraph. This method is illustrated in Fig.~\ref{fig:subgraphs}, where (a) shows the original particle-patch graph and (b) the pruned cyclic subgraph.

We call the subgraph of the particle-patch graph induced by all nodes not contained in the cyclic subgraph the acyclic subgraph. It contains only structural components which correspond to linear or branched chains within the particle network. These are either isolated or dangling from the network of loops.

\subsection{Loop Detection}
For the structural characterization of two-dimensional porous networks, we focus on the detection of loops which we identify as the boundaries of the pores. 

We find the loops within the porous network by analysis of the cyclic subgraph, as introduced in Sec.~\ref{sub:subgraphs}. Fig.~\ref{fig:subgraphs}(b) shows the cyclic subgraph of a kagome nucleus as an embedding in the plane, i.e. a drawing of the graph on that surface without overlapping edges.

For two-dimensional particle networks, the relative configuration of the particles directly provides such an embedding of the graph. An embedding is specified by a so-called rotation system, i.e. the cyclic ordering of edges incident to each node in the graph. We choose the cyclic ordering according to the bond orientations within the particle configuration. For attaining the rotation system, we record at each node the ordering of its edges, according to the particle orientation and the orientation of the incident bonds. This defines the embedding that naturally reproduces the pore structure of the particle network. This embedding allows the loops to be identified as faces of the cyclic subgraph. Here, a face is one of the regions into which the embedding divides the surface.

For detection of the faces we apply the face-walking algorithm \cite{FaceWalk,FaceWalkBoostLibrary}. As input the algorithm takes the rotation system, using it to guide walks around closed paths circling the faces of the graph. The face-walking algorithm returns all face cycles of the cyclic subgraph, from which the complete set of loops $\mathcal{L}$ can be derived. For further details we refer to the Appendix.

We note that the presented approach of loop detection is methodologically distinct from established graph-based methods for ring detection, as applied for example in the analysis of silica glasses \cite{Lichtenstein2012,Buchner2017,AltmanMalashevich2016,Gao2016}. Most commonly applied algorithms detect primitive rings, where the primitive ring is defined as a cycle in the graph that cannot be decomposed into a number of smaller cycles. To detect such primitive rings, the algorithms rely on the shortest-path method \cite{Cormack2002,Goetzke1991}.

\subsection{Loop Order Parameters}\label{sec:loop}
\begin{figure}
    \centering
    \includegraphics[width=0.5\linewidth]{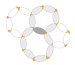}
    \caption{A particle (dark-gray) has at most two adjacent loops (light-gray), which define its local neighborhood for the local loop order parameter in Equ.~(\ref{equ:locloopOP}).}
    \label{fig:locality}
\end{figure}

The knowledge of the complete set of loops $\mathcal{L}$ allows us to formulate a novel local- and global loop order parameter, $\gamma_{6}(i)$ and $\Gamma_{6}$. We construct the local order parameter $\gamma_6(i)$, defined for a particle $i$, as a local measure of the network regularity in terms of 6-loops, i.e. loops of size six. For locality we use particle $i$'s immediate neighborhood, defined as the at most two adjacent loops to the particle, as illustrated in Fig.~\ref{fig:locality}. We describe loops as a sequence of particle indices, and denoting these sequences as $L_{i,l}$ of cardinality $|L_{il}|$, with the loop index $l\le2$. By assigning a weight factor $w_6(|L_{i,l}|)\in[0,1]$ to each adjacent loop, we define the local loop order parameter as:
\begin{align}\label{equ:locloopOP}
    \gamma_6(i) = \frac{1}{2}\sum_{l\in l_i}w_6(|L_{i,l}|)\text{ .}
\end{align}

We denote by $l_i$ the index set of the loops adjacent to particle $i$:
\begin{align}
l_i = \begin{cases}
\{1,2\} & \text{if }i \text{ is adjacent to two loops}  \\
\{1\}   & \text{if }i \text{ is adjacent to one loop}  \\
\{\}    & \text{if }i \text{ is adjacent to no loop}
\end{cases}
\end{align}

We choose a symmetric, Gaussian weight depending on how close the loop size $\ell =|L_{i,l}|$ is to the target loop size $\ell_0=6$,
\begin{align}\label{equ:locloopOP2}
    w_{\ell_0}(|L_{i,l}|) = \exp \Bigg[\frac{-(|L_{i,l}|-\ell_0)^2}{\Delta}\Bigg] \text{,}
\end{align}
with $\Delta=9$, resulting in weights given as example for 7- up to 11-loops: $w_6(7)\approx0.895$, $w_6(8)\approx0.641$,  $w_6(9)\approx0.368$,  $w_6(10)\approx0.169$, and $w_6(11)\approx0.062$.

For the global loop order parameter $\Gamma_{6}$ we go from a particle-based to a loop-based descriptor, covering all $M$ loops in the system in one parameter. Here we denote a loop in the system as $L_l$, with the loop index $l=1,...,M$ covering all loops in the system. We define the global descriptor by taking a weighted sum over all loops, where we adopt again the Gaussian weight factor $w_6(L_l)$, now for each loop $L_l$, giving full weight to 6-loops and a smaller weight to all other loop sizes. For normalization we take the number of loops that would appear in a bulk regular loop network of 6-loops, $M^{\text{reg-}6}$ (detailed approximation in the Electronic Supplementary Information (ESI)). In total, the global loop order parameter is then defined as:
\begin{align}\label{equ:loopOP}
     \Gamma_6 = \frac{1}{M^{\text{reg-}6}}\sum_{l=1}^{M}w_6(L_{l})\text{.}
\end{align}

More general loop order parameters $\Gamma_{\ell_0}$ and $\gamma_{\ell_0}(i)$, targeting different loop sizes $\ell_0$ can be defined in a similar manner.

\section{Results}\label{sec:results}
\begin{figure}
    \centering
    \includegraphics{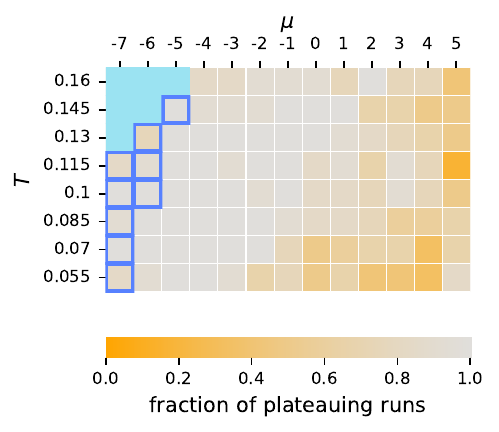}
    \caption{
    State diagram in the plane of chemical potential $\mu$ and temperature $T$. Light blue indicates the gas phase, mid-blue frame indicates nucleating states, all other states enter the production run in an assembled state. Color scale of the nucleating and assembled states denotes the fraction of trajectories that reach an energy plateau within the simulation time, ranging from 0 (orange), where none of the trajectories plateau, to 1 (light gray), where all trajectories do.}
    \label{fig:equ_rates}
\end{figure}

We study the two-dimensional self-assembly of colloidal ellipses with two patches at opposite poles over a range of chemical potentials \(\mu\in[-7,5]\) and temperatures \(T\in[0.055,0.16]\) (see Sec.~\ref{sec:methodsandmodels} for model and simulation details). We immediately expose the dilute initial state to the target thermodynamic condition, temperature $T$ and chemical potential $\mu$, and follow its subsequent relaxation and self-assembly. With this simulation protocol we observe distinct characteristics of the total energy trajectory depending on the thermodynamic state points, which we analyze as proposed in Section~\ref{sec:methodsandmodels}.

Based on the time evolution of the total energy, we distinguish three qualitative self-assembly scenarios: (1) the system remains in the dilute gas phase; (2) a network forms by nucleation during the production run; and (3) assembly occurs rapidly, such that the system enters the production run in an assembled state and subsequently relaxes toward an equilibrated state. In the latter scenario, the relaxation rate varies considerably across thermodynamic conditions and trajectories, and not all trajectories reach an energy plateau within the simulation time.

Figure~\ref{fig:equ_rates} summarizes these behaviors in the $\mu T-$ plane. Gaseous states (light blue) occupy the high-$T$, low-$\mu$ corner of the diagram, where assembly is least favorable. At low chemical potential, lowering $T$ first brings the system into a nucleating regime (outlined in mid-blue), where assembly proceeds by nucleation during the production run. At still lower $T$, the system enters a regime in which assembly already occurs before the production run begins. Increasing $\mu$ produces the same progression albeit starting at higher $T$ : at a given $T$, the system crosses from gaseous to nucleating to rapidly assembling states. For the nucleating and assembled states, the heat map color denotes the fraction of trajectories that reach an energy plateau within the simulation time, ranging from orange (0, where none plateau) to light gray (1, where all plateau). Consequently, the crossovers between these regimes form a roughly diagonal pattern in the $\mu T-$ plane, reflecting the combined effect of temperature and chemical potential on the tendency to assemble.

Within the assembled region, the fraction of plateauing trajectories decreases continuously towards the low-\(T\), high-\(\mu\) corner. This is consistent with the increasingly deep quench into the assembled regime, where increased bond stability (at low $T$) and higher densities (at high $\mu$) make structural rearrangements progressively slower. Consequently, fewer trajectories reach an energy plateau within the accessible simulation time.
We next characterize the structures formed across the \(\mu T-\)plane. Inspection of the full parameter space reveals systematic structural trends with both chemical potential and temperature. We therefore will focus on two representative, orthogonal cross-sections that capture these trends: one at fixed temperature $T=0.145$, along which we vary \(\mu\), and one at fixed chemical potential $\mu=1$, varying \(T\). We conclude this section by separately analyzing three temperatures at \(\mu=-7\) for which nucleation occurs during the production run.

A more extensive analysis across other intersections in the \(\mu T-\) plane is provided in the Electronic Supplementary Information (ESI). 

\subsection{Analysis of assembled network structures at fixed temperature}\label{sub:fix_temp}
\begin{figure*}[t]
    \centering
    \includegraphics[width=\textwidth]{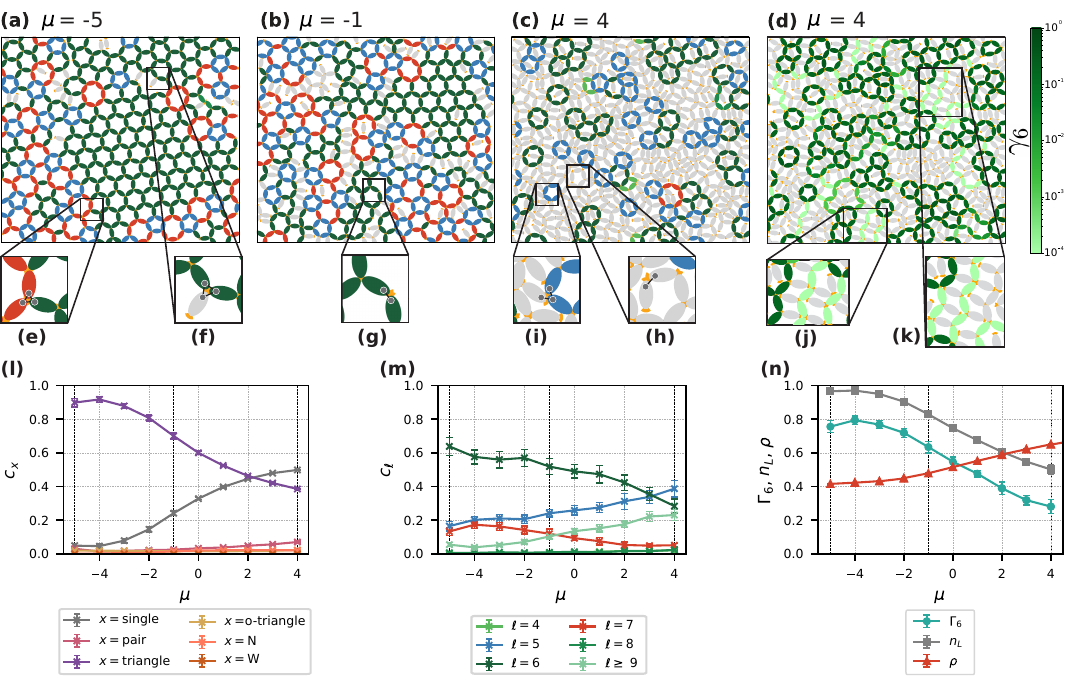}
    \caption{Panels (a)-(c): snapshots of configurations at temperature $T=0.145$ with chemical potential at (a) $\mu=-5$; (b) $\mu=-1$; (c) $\mu=4$. Particles are colored according to their loop size: particles in 4-loops (light-green), 5-loops (blue), 6-loops (dark-green), 7-loops (red) and 8-loops (light-green). Particles not in a loop or in a loop of size $\ge9$ are colored in light-gray. As a particle may belong to up to two loops, we visually resolve this ambiguity by picking the color according to this order: 5-loops, 7-loops, 6-loops, 4-loops, 8-loops. This hierarchy gives priority to 5- and 7-loops enhancing the visibility of these defects. Panel (d) shows the same configuration as panel (c) with the particles colored according to the local loop order parameter $\gamma_6$ (see Equ.~\ref{equ:locloopOP}), with the color gradient following a logarithmic scale to enhance the visibility of the larger loops.
    Panels (e)-(i) exemplify the occurrence of bond motifs within the network: (e) a triangle motif within the kagome lattice; (f) a triangle motif at the boundary of the crystalline zone; (g) a pair motif within a loop; (h) a pair motif as part of a particle chain; (i) an open-triangle motif.
    Panels (j-k) show non-looped, regular structures that are potentially crystalline and in competition with the kagome lattice.
    Panel (l) shows the average bond-motif concentration $\bar{c}_x$ for bond motifs of type $x$; panel (m) the average concentration of $\ell$-loops, $\bar{c}_{\ell}$; and panel (n) global loop order parameter $\Gamma_6$, the average number fraction of particles within loops $\bar{n}_L$ and the reduced density $\rho$. Only runs whose energy trajectory reached a plateau (criterion in Sec.~\ref{sub:plateau}) are included. Each run is time-averaged over its plateau region, and symbols show the mean of these run-averages over the independent runs. Error bars show the standard deviation between runs.}
    \label{fig:mu_analysis}
\end{figure*}

For the first cross-sectional analysis we use $T=0.145$ as it gives us a full overview of different assembly scenarios as observed in the state diagram in Fig.~\ref{fig:equ_rates}: the gas state at $\mu=-7$ to $-6$ (light blue), a nucleating state point at $\mu=-5$, initially assembled and equilibrating states at $\mu=-4$ to $1$, and states that do not equilibrate within the course of our simulation, and rather exhibit a creeping behavior with the energy still slowly decreasing at $\mu=2$ to $5$.

In Fig.~\ref{fig:mu_analysis}(a-d) we show snapshots of different $\mu$-values at $T=0.145$. Generally, across all displayed $\mu$-values we observe porous assembled networks.
At low chemical potential, at $\mu=-5$ these networks are crystalline with a kagome order that has 6-loops as the dominant local structural motif (see panel (a) of Figure~\ref{fig:mu_analysis}). Highlighting these crystalline domains, particles within a 6-loop are colored in dark green. We observe that the large crystalline domain in the center, is interrupted by more defect-rich regions where 5-loops (blue) and 7-loops (red) are predominant. We note 4-loops and loops with a size larger $\ge8$ (both denoted in light green) are very rare, as are particles that do not belong to any loop (in gray).

As we move to higher chemical potentials at $\mu=-1$ (panel (b) in Fig.~\ref{fig:mu_analysis}), the crystalline order starts disappearing. We find domains of crystalline 6-loops, but with their size considerably decreased. While 5-7 defects are still present, we see more grain boundaries with particles that are not in loops. At positive chemical potential, $\mu=4$ (panel (c) in Fig.~\ref{fig:mu_analysis}), domains of 5-,6- and 7-loops are diminished dramatically, giving way to a disordered but still porous network. Besides non-looped particles, we also observe a few loops with sizes larger than $8$, visible in panel (d) of Fig.~\ref{fig:mu_analysis}. The same snapshot as panel (c) is shown here, but with particles colored according to their loop order parameter, as defined in Eq.~\ref{equ:locloopOP}.

The looped and non-looped structural motifs are built from a small set of local patch-bonding motifs. Here we focus on the three most prevalent ones: pairs, triangles, and open triangles, all of which can be identified within the particle-patch network as described in \cref{sub:bondingmotifs}. For reference, these motifs are highlighted in the insets of Fig.~\ref{fig:mu_analysis}(e--h). Triangles form the basic building block of connected loops (panel (e)), but can also occur outside loops, either isolated or along chains (panel (f)). Pair bonds (panels (g) and (h)), and open triangles (panel (i)), likewise occur both within and outside looped structures.

In panel (l) of Fig.~\ref{fig:mu_analysis} we show the concentrations of these bonding motifs $\overline{c}_{x}$, i.e. the fraction of bonding motifs of type $x$ over the number of all bonding motifs appearing as patch clusters in the particle-patch graph (see Fig.~\ref{fig:graph} for an illustration of these motifs). We find that, across all chemical potentials triangles (in purple) are the most prevalent bonding motif, where the fraction of triangles is close to $1$ at the lowest $\mu$-values and then monotonically decreases towards positive $\mu$-values. At the same time the fractions of pairs (pink), negligible at $\mu=-5$, increases monotonically. The open triangle motif (yellow) and the N-motif (light orange) increase slightly, while singles (gray) and W-motifs (dark orange) remain negligible across all $\mu$-values. We note that the bonding motif distribution follows qualitatively similar trends across other chemical potential cross-sections at constant $T$ (see the Electronic Supplementary Information (ESI)).

The decrease in triangle motifs with increasing chemical potential is also consistent with the representative snapshots, where we have identified an overall loss of loop network domain size, in which the triangle motif appears as the basic local building block. To quantify this trend directly, in Fig.~\ref{fig:mu_analysis}(m) we plot the fraction of loops of different sizes as a function of chemical potential. Here, the number of loops of size \(\ell\) is defined relative to the total number of loops \(M\) in the system as $c_{\ell}=\frac{M_{\ell}}{M}$.

At the lowest chemical potential, \(\mu=-5\), 6-loops clearly dominate, with \(c_6\approx0.6\). Their fraction then decreases steadily with increasing \(\mu\), reaching slightly below \(0.4\) at the highest chemical potential. The 5- and 7-loops both start at around \(0.2\), but then split: 5-loops become more frequent, while 7-loops continuously disappear and are almost absent at the highest \(\mu\). At the same time, larger loops with \(\ell\geq8\), which are almost absent at low \(\mu\), become increasingly important and reach a fraction of about \(0.2\) at $\mu = 4$.

The loop statistics thus show more than just a loss of 6-loops. With increasing chemical potential, the loop-size distribution becomes broader and loses the balance between 5- and 7-loops that is present at low \(\mu\). This balance is expected in (a predominantly) kagome order, where 5- and 7-loops appear as paired topological defects of the hexagonal lattice \cite{Mermin1979}. Once this hexagonal order of the 6-loops breaks up, the separation of the 5-7 concentrations reveals a clear preference for 5-loops. We observe that this shift from balanced to unbalanced 5- and 7-loop concentrations with increasing chemical potential is persistent across all temperatures, with increasing imbalance at higher chemical potentials (see the Electronic Supplementary Information (ESI)).

This imbalance of 5- and 7-loop concentrations is in line with ring characteristics in trivalent patchy particles as observed in Ref.~\cite{SwinkelsSchall2023}, which leads us to speculate that 5-loops have higher kinetic accessibility compared to 7-loops.

The relative loop concentrations, however, only tell us what kinds of loops are present. They do not tell us how much of the system is actually organized in loops, or how much 6-loop order survives on the system level. For this reason, we proceed by plotting the fraction of particles belonging to loops, \(n_L=N_L/N\), together with the global 6-loop order parameter \(\Gamma_6\), defined in \cref{equ:loopOP}, and the density \(\rho\) in Fig.~\ref{fig:mu_analysis}(n). At low \(\mu = -5\), with \(n_L\approx1\), almost all particles are within loops, indicating a fully connected loop network. The few particles outside loops are visible in snapshot (a), mainly around defects and grain boundaries. At the same time, \(\Gamma_6\) is high and reaches values around \(0.8\), reflecting the dominance of the 6-loop structure, but with some defect loops of different size. As \(\mu\) increases, the density rises, while both \(n_L\) and \(\Gamma_6\) decrease. The system becomes denser, and at the same time less loop-dominated and less ordered with respect to the 6-loop structure.

With the gradual loss of the 6-loop structure at increasing density, other higher density, non-looped structures are introduced, as can be seen in panel (d). We visually identify some of these structures (see Fig.~\ref{fig:mu_analysis}(j-k)) as locally ordered, built from interlaced chains of triangles, giving them the appearance of a zig-zag ladder. These structures could potentially appear as a competing crystalline structure with a higher density compared to that of the kagome lattice. At the same time the local density in kagome lattice regions is also increased by a gas of particles occupying the pores of the network. We stress that these observations are speculative, as a detailed analysis of free energies is beyond the scope of this study.

\subsection{Analysis of assembled network structures at fixed chemical potential}\label{sub:fix_chem}
\begin{figure*}[t]
    \centering
    \includegraphics[width=\textwidth]{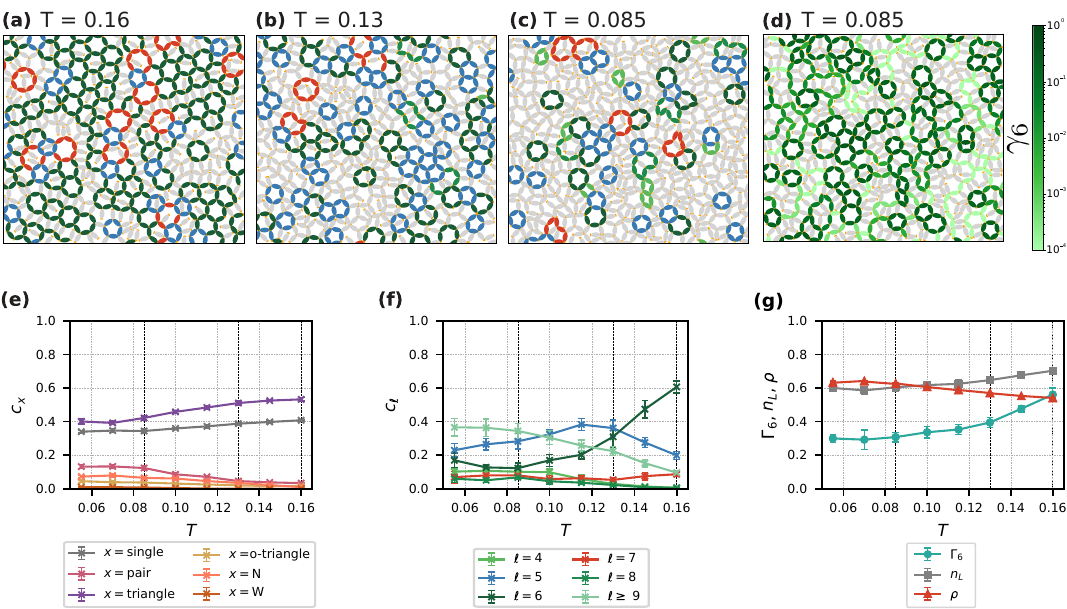}
    \caption{Panels (a)-(c): snapshots of configurations at chemical potential $\mu=1$ with temperature at (a) $T=0.16$; (b) $T=0.13$; (c) $T=0.085$. Particles are colored according to their loop size: particles in 4-loops (light-green), 5-loops (blue), 6-loops (dark-green), 7-loops (red) and 8-loops (mid-green). Particles not in a loop or in a loop of size $\ge9$ are colored in light-gray. As a particle may belong to up to two loops, we visually resolve this ambiguity by picking the color according to this order: 5-loops, 7-loops, 6-loops, 4-loops, 8-loops. This hierarchy gives priority to 5- and 7-loops enhancing the visibility of these defects. Panel (d) shows the same configuration as panel (c) but with the particles colored according to the local loop order parameter $\gamma_6$. Panel (e) shows the average bond-motif concentration $\bar{c}_x$ for bond motifs of type $x$; panel (f) shows the average concentration of $\ell$-loops, $\bar{c}_{\ell}$. In panel (g) the global loop order parameter $\Gamma_6$, the average number fraction of particles within loops $\bar{n}_L$ and the reduced density $\rho$ are shown. Only runs whose energy trajectory reached a plateau (criterion in Sec.~\ref{sub:plateau}) are included. Each run is time-averaged over its plateau region, and symbols show the mean of these run-averages over the independent runs. Error bars show the standard deviation between runs.}
    \label{fig:T_analysis}
\end{figure*}

We continue our analysis by inspecting the $(\mu,T)$-cross-section at constant $\mu=1.0$, shown in Fig.~\ref{fig:T_analysis}. In panels (a)-(c) we display selected snapshots for the temperatures (a) $T=0.16$ , (b) $T=0.13$ and (c) $T=0.085$. Generally we observe porous networks with varying degrees of crystalline order that is dependent on temperature. At the highest temperature $T=0.16$, we find the predominant local motif to be 6-loops, assembled into kagome crystal domains. We find these domains to be separated by grain boundary defects, and interrupted by defect loops of size 5 and 7, where they tend to appear in close proximity to each other forming pairwise 5-7 defects.

At $T=0.13$, in panel (b), the number of 6-loops seem to have diminished significantly while in this snapshot 5-loops appear as the most dominant loop structure. While the kagome lattice order is lost, 5-, 6- and 7-loops still do cluster together forming local domains of disordered loop structures. The overall number of 5-,6- and 7-loops appears diminished in favor of either larger loops or non-looped motifs. 

For the lowest temperature $T=0.085$, shown in panel (c), this trend continues: the number of 5-, 6-, and 7-loops seems reduced further in favor of larger loops or non-looped motifs. To determine if larger loops are present for this lowest temperature, in panel (d) we show the same snapshot as in (c) with particles colored according to the loop order parameter. While we do find that this system has a significant amount of loops larger than 8, as indicated by the lighter shades of green, non-looped structures appearing in gray are also present in high numbers.

The bonding motif concentrations (panel (e)) shows the triangle motif as monotonously increasing with $T$ from $\approx 0.4$ at $T=0.055$ to above $0.5$ for $T=0.16$. The concentration of open patches, or singles, increases parallel to the triangle motif. This increase of triangle and single motifs occurs at cost of all other motifs. For low temperatures the pair, the N- and open-triangle motif are all present in significant amounts, but at high temperatures these bonding motifs almost completely vanish.
Overall the quantitative analysis of the bonding motifs confirms the observations based on the selected snapshots: At high $T$ triangle bonding motifs are dominant as they form the basis for the prevalent 5-, 6-, and 7-loops. At the same time the single motif occurs at significant concentration, either occupying the pores of the kagome lattice or present as dangling particles at the grain boundaries.

To complement this analysis, panel (f) displays the distribution of loop sizes. We do see a rapid increase in the concentration of 6-loops with rising temperature from below $0.2$ at $T=0.055$ to $0.6$ at $T=0.16$, where 6-loops are dominant over all other loop types. 
While the 6-loops show a mostly monotonic increase, the 5- and 7-loops behave differently: The 5-loops show a non-monotonic behavior by starting as the dominant loop type at $\approx 0.2$ for the lowest $T$, then increasing to a maximum to $\approx 0.4$ at $T=0.115$ and finally dropping again to $0.2$ at $T=0.16$; the 7-loops remain approximately constant at $\approx 0.1$ across all $T$. Larger loops (\(\ell\ge9\)) are most prevalent at low temperature, where they are the most dominant class of loops with a concentration of $\approx 0.4$. As the temperature increases, larger loops diminish to $\approx 0.1$. At $T=0.085$ we see by comparison of Fig.~\ref{fig:T_analysis}(c) and (d) that these longer loops (in lighter shades of green) arise in very irregular arrangements, connecting the clusters of smaller loops, and interpenetrating regions of particles organized in non-looped structures.

To compare the percentage of loops to the occurrence of other, non-looped mesoscopic motifs, we plot the fraction of particles in loops $\bar{n}_{\rm L}$ in panel (g) of Figure~\ref{fig:T_analysis}. Starting at a value of $\bar{n}_{\rm L}\approx 0.6$ for the lowest temperature, we observe that $\bar{n}_{\rm L}$ only slightly increases to $\bar{n}_{\rm L} \approx 0.7$ at $T=0.16$. Compared to the significant change in the loop size distribution towards more 6-loops, this observation may be surprising. However, at high temperatures the loop concentration may be lower than expected due to a large concentration of single motifs, appearing as a particle gas in kagome lattice pores as well as at grain boundaries -- a fact that is also reflected in the relatively constant density $\rho$ across temperatures. At low temperatures, in turn, the number fraction of loops may be higher than expected due to the increase of large loops (see panel (d) and (f)). 

The observation that kagome order together with the occurrence of 6-loops increases with temperature is therefore better reflected in the loop order parameter $\Gamma_6$ in panel (g), that is below $<0.4$ for temperatures $T\le0.13$ and then increases to $\approx0.6$ for the highest temperature.

We conclude by an overall comparison of the cross-section at constant $\mu=1.0$, see Fig.~\ref{fig:T_analysis}, to the cross-section at constant $T=0.145$, in Fig.~\ref{fig:mu_analysis}. 
In both cross-sections we find a structural change from ordered states of predominantly 6-loops, presenting in a kagome lattice (at high $T$ and low $\mu$), towards more disordered states, with significantly lower concentrations of 6-loops (at low $T$ and high $\mu$). With the loss of the ordered structures, the loop size distributions broaden, showing a general increase in larger loops ($\ell\ge9$) as well as a growing imbalance between 5- and 7-loops, with 5-loops as the preferred motif. While at lower temperatures, and with it higher bond stability, we speculate that the 5-loops are kinetically more accessible, higher concentrations of 5-loops at high chemical potentials may be due to optimizing local packing in the disordered networks.

Tracking the bonding motifs in direction of decreasing structural order, we find that in both cross-sections (Fig.~\ref{fig:mu_analysis}(l) at constant $T=0.145$ and Fig.~\ref{fig:T_analysis}(e) at constant $\mu=1.0$) all but the single bonding motif play a qualitatively similar role: while in the cross-section at constant $T=0.145$ the ordered state includes almost exclusively triangle motifs, the ordered state in the cross-section at constant $\mu=1.0$ includes almost the same concentration of single and triangle motifs.  We attribute this difference to increased occurrence of grain boundaries and the particle gas occupying pores for higher chemical potential. 

\begin{figure*}[t]
    \centering
    \includegraphics[width= \textwidth]{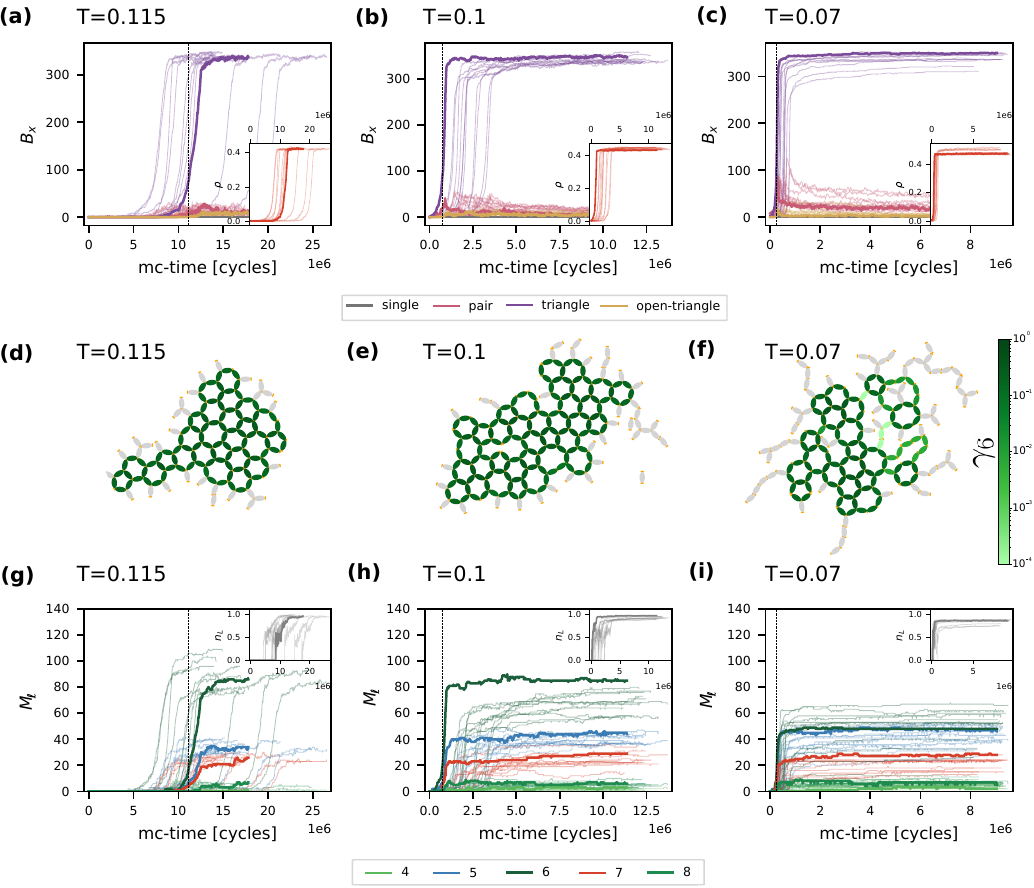}
    \caption{Panels (a)-(c): bonding motifs $x$ in absolute counts $B_{x}$ and density $\rho$ (insets) of all twelve trajectories as function of mc-time at chemical potential $\mu=-7$ for temperatures (a) T=0.115, (b) T=0.1, (c) T=0.07. Colors: singles (gray), pairs (pink), triangles (purple), open-triangles (orange). For comparison we highlight one set of curves that belong to one selected trajectory. Panels (d)-(f): Snapshots of nuclei of the highlighted trajectories at mc-times denoted by dotted vertical gray lines in (a)-(c) at (d) T=0.115, (e) T=0.1, (f) T=0.07. Particles in loops colored according to local loop order parameter $\gamma_6$; particles not in any loop denoted in gray. Panels (g)-(i): Number of loops $M_{\ell}$ as function of mc-time for loop sizes $\ell=4$ (light green) $\ell=5$ (blue), $\ell=6$ (dark green), $\ell=7$ (red) and $\ell=8$ (mid-green) for (g) $T=0.115$, (h) $T=0.1$,  (i) $T=0.07$. Insets in (g)-(i): fraction of particles in loops $\bar{n}_L$ (gray) as function of mc-time.}
    \label{fig:nucl-processes}
\end{figure*}

\subsection{Self-assembly via nucleation}\label{sub:nucleation}
In the range of negative chemical potential, between $\mu=-7$ and $\mu=-5$, for some states we observe the full assembly processes during the production run. We identified this process as nucleation into a kagome crystal.
These states directly nucleate from vapor and equilibrate after completing the assembly process according to our energy plateauing criteria and are encircled by a mid-blue frame in Fig.~\ref{fig:equ_rates}.\medskip

In Fig.~\ref{fig:nucl-processes} we first analyze the process macroscopically by tracking the density as a function of mc-time, as shown in insets of panels (a)-(c), followed by a study of the microscopic pathway
via bonding motifs, shown in the main panels (a)-(c), snapshots of nuclei in (d)-(f), and loop statistics in (g)-(i).

This detailed analysis of the assembly pathway gives us the opportunity to compare patchy ellipses with their spherical counterparts -- the triblock patchy particles, where the two patches sit on opposite poles and are large enough to host two bonds per patch.  
In systems of spherical triblock particles nucleation proceeds via a two-step process, where the particles first assemble into chains and only subsequently, these chains assemble into a kagome lattice structure \cite{GranickChen2011, Romano2011, Bahri2022}. At this point it is interesting to note that this two-step pathway has been observed in experiments \cite{GranickChen2011} as well as in various independent simulation studies with varying types of interaction potentials \cite{Romano2011, Bahri2022, Gharibi2024}. With this study of patchy ellipses, we can pose the question: how the assembly pathway is affected if the particles become elongated in one direction? 
In an attempt to answer this question, we start at the macroscopic analysis by tracking the density across mc-time in the insets of ~\ref{fig:nucl-processes}(a)-(c) at $\mu=-7$ for three different temperatures $T=0.115$, $T=0.1$ and $T=0.07$ and all parallel runs for each state point.

All three density plots show behavior expected of a nucleation scenario. The densities initially remain very low, consistent with the vapor state, followed by a sharp increase in density whose onset varies between individual trajectories. At later times, the density levels off into a high-density plateau, indicating that the system has completed its assembly into the higher density state. 
We observe that from the highest temperature $T=0.115$ to the lowest $T=0.07$ the nucleation-time distribution shifts strongly towards shorter times. This trend is consistent nucleation behavior, as an increasing temperature quench increases the nucleation rate.
We note that the density jumps in a single step for all temperatures. While this does not yet indicate a single structural nucleation scenario, it excludes the possibility of metastable intermediaries with a differing density. 

At this point, we must emphasize that the present analysis of the nucleation pathways remains qualitative. With only twelve independent trajectories per state point, the statistics are insufficient for a quantitative characterization of the nucleation process. Determining quantities such as the nucleation rate, critical nucleus size, or free-energy barrier would require extensive trajectory sampling, which lies beyond the scope of this study.

In the next step we analyze the microscopic structure by tracking bonding motifs over mc-time in the main panels of Figure~\ref{fig:nucl-processes}(a)-(c), according to section~\ref{sub:bondingmotifs}. We record singles (in gray), pairs (in pink), triangles (in purple) and open-triangles (in orange) in absolute counts. While we do plot the results for all twelve runs per state point, for direct comparison we have highlighted the bonding motif set for only one run.

As pairs are the bonding motif at the basis of chain formation, and triangles are the basic motif of the kagome lattice, in a two step nucleation process that involves intermediary chain formation we would expect the pairs to increase before and independently of the triangles. Comparing the number of identified pairs and triangles for all three temperatures (Fig.~\ref{fig:nucl-processes}(a)-(c)), we find that, the pair and triangle counts increase approximately at the same time, where the triangles outgrow the pairs significantly soon after the onset of nucleation. We therefore conclude that there is no evidence for a pronounced chain-dominated intermediate state preceding kagome ordering as explicitly observed in Ref.~\cite{Gharibi2024} for spherical triblock patchy particles. Instead, triangle-rich structures emerge together with the nucleating aggregate.
We additionally confirm that nucleation proceeds via the growth of the largest cluster in the system for all observed temperatures and supply a plot of growth of the largest bonded cluster in the Electronic Supplementary Information (ESI). 

In a detailed comparison of all three temperatures we note that the triangles tend to increase monotonically after the onset of nucleation and reach a maximum number of $\approx 330-350$ after equilibration, the pair bonds tend to peak during the nucleation event and then decrease towards equilibration.
For the highest temperature $T=0.115$ this peak reaches $\approx 10$, for $T=0.1$ it reaches $\approx 30-50$ and for the lowest temperature $T=0.07$ the pair count reaches up to $80-100$. To understand this temperature dependent behavior, in Fig.~\ref{fig:nucl-processes}(d)-(f) we show nuclei snapshots of the highlighted curves around the onset of nucleation, indicated by the vertical dotted lines in Fig.~\ref{fig:nucl-processes}(a)-(c) respectively. The particles in the nuclei are colored according to the local loop order parameter $\gamma_6$, where dark green, indicates a high degree of 6-loops, while for more light green shades, the loop size deviates significantly from six particles.
For all three temperatures we observe that pair bonded particles are predominantly located at the periphery of the nuclei and most of them do not contribute to loops (gray) or are part of significantly larger loops than size six (light-green).

We observe that at the highest temperature, $T=0.115$, the nucleus is compact and displays high local kagome order that coincides with a high local loop order parameter $\gamma_6$ as is highlighted by the dark green coloring of the particles, and has a relatively sharp interface. As the temperature decreases, at $T=0.1$, we observe that the kagome nucleus has some chain-like appendages and a noticeably rougher perimeter. At the lowest temperature shown, $T=0.07$, the shown nucleus is considerably more diffuse and ramified, exhibits substantially less regular loops, and develops extended chain-like protrusions from the central aggregate. We therefore conclude that the increased pair count at decreasing temperature is indicative of a lower crystalline order and rougher boundaries. Moving from the pair and triangle counts, which remain indirect measures of local order, in panels (g)-(i) of Figure~\ref{fig:nucl-processes} we explicitly plot the time evolution of absolute number of loops of different sizes, $L_{\ell}$, as well as the fraction of particles within loops (gray curves in the insets) $\bar{n}_{L}$. These plots support the analysis drawn from the snapshots and the bonding motifs: the number of 5-,6-, and 7-loops as well as $\bar{n}_{L}$ rise sharply and monotonically with the onset of nucleation for all observed temperatures before reaching a plateau at the assembled phase. Taken together, these plots support the conclusion that the lattice forms immediately without a chain intermediary, and with the most prevalent loop sizes being 5-, 6- or 7-loops. Inspecting the differences between the temperatures across trajectories, we note that the maximum number of 6-loops relative to the numbers of all other loop sizes is on average highest for $T=0.115$, followed by $T=0.1$, where on average it slightly decreases, and it is lowest for $T=0.07$. 
Maximum numbers of 5-loops and 7-loops remain lower than the number of 6-loops for $T=0.115$ and $T=0.1$. At $T=0.07$ the number of 6-loops decreases significantly: in fact for the example trajectory in bold, there is an equal number of 5- and 6-loops ($\approx 50$). 
As observed in Sec.~\ref{sub:fix_temp} and ~\ref{sub:fix_chem}, the numbers of 5- and 7-loops are imbalanced, with 5-loops appearing in larger numbers throughout the nucleation process, as also illustrated by example curves in Fig.~\ref{fig:nucl-processes}(g). This quantitative analysis over all trajectories corresponds well with the snapshots: for $T=0.115$ (see Fig.~\ref{fig:nucl-processes}(d)) and $T=0.1$ (see Fig.~\ref{fig:nucl-processes}(e)) the predominant loops are 6-loops, with some 5- and 7-loops in between, while for $T=0.07$, the number of 6-loops is decreased in favor of 5-loops.\medskip

Taken together, these observations suggest that nucleation in the patchy ellipses does not proceed through a well-separated chain intermediate. Instead, locally kagome-like loop structures form immediately. Lowering the temperature accelerates this process but simultaneously promotes chain-like growth and defective loop structures, producing increasingly ramified and less 6-loop-selective nuclei.

\section{Conclusion \& Discussion}\label{sec:conclusio}
In this work, we studied the assembly of patchy elliptic particles, that can be viewed as an elongated extension of spherical triblock particles \cite{Romano2011}, into porous network structures. 
By example of two cross-sections on the $\mu T-$plane, one at constant $T=0.145$ and another at constant $\mu=1.0$, we have shown how the loop network structure changes from ordered to disordered states. Ordered network states appear as the kagome lattice, which is to the best of our knowledge, the first observation of patchy ellipses assembly into this structure. We have analyzed this order-disorder transition in terms of bond motif type and loop size distributions, as well as the loop order parameter. The loss of this loop order picks up with increasing chemical potential and decreasing temperature. With the increasing loss of the kagome order, we identified the following phenomena: As a common motif, we see that 5- and 7-loop concentrations become imbalanced, with the 5-loops appearing as the preferred motif. At the same time, the concentration of larger loops increases and that of smaller loop sizes decrease. A similar imbalance in prevalence of 5- and 7-loops has been observed in trivalent patchy particles in Ref.~\cite{SwinkelsNoya2023,SwinkelsSchall2023}. Together with this shift in the loop size distribution towards larger loop sizes, we also see other, non-looped structures appearing. Among these non-looped structures, we visually identified a regular structure, that we hypothesize could be in competition with the kagome lattice.

In addition to the structural analysis of the assembled states, we have followed the emergence of the particle networks from the dilute phase. For very low chemical potentials, we find that assembly into the kagome structure proceeds via nucleation. However, in contrast to the two-step nucleation process observed for spherical triblock patchy particles \cite{GranickChen2011, Romano2011, Bahri2022, Gharibi2024}, we observe a nucleation proceeding directly from the particle gas. As the patchy ellipses studied in this contribution can be viewed as an elongated version of the triblock patchy spheres, this leads us directly to speculate about the role that particle shape plays in these nucleation scenarios: we hypothesize that the differences in phenomena originates largely from a higher bond-bending flexibility of the triangle bonding motifs compared to that of the patchy triblock spheres. With a higher flexibility, the bonding volume and thus the entropy is enhanced. This changes the kagome lattices overall stiffness and by this its ability to host defect loop sizes, such as 5-7 defect pairs. Specifically the changed entropy directly influences the relative stabilities of the equilibrium phases, and therefore partly determining the pathway during nucleation.

With regards to the structural analysis of the assembled particle networks, the main methodological innovation of this contribution is the introduction of the particle-patch graph. Contrary to graph representation abstracting particles as single nodes, the particle-patch graph is able to distinguish between the bonds being formed at different patch sites. This allows us to structurally distinguish local bonding motifs, that are missed by standard graph representations. We introduced methods, applied on the particle-patch graph, that detect all bonding motifs and all loops directly, where the loops are defined as the boundaries of the pores, subdividing the plane into distinct pore regions.

Our observations of highly disordered network structures demonstrate the need for a graph-theoretical treatment, and the methods presented here provide a well-suited basis for further analysis aimed at a full characterization of these networks. Beyond the identification of loops, that we have presented in this contribution, such a characterization includes the identification of regular, non-looped structures that emerge in the acyclic subgraph as potential competing crystal structures. Zooming out to larger scales, the spatial arrangement and connectivity of looped structures, as well as of other mesoscopic motifs, are a natural direction to pursue. For this purpose, a graph representation that abstracts loops as nodes and connects neighboring loops by edges will be a valuable tool.

With this we conclude, that the presented particle-patch graph allows for a full particle-based structural characterization, serves as the starting point for a hierarchy of graph abstractions, connecting the spatial scales from local bonding motifs, via mesoscale loop structures up to connected loop networks.

\section*{Conflicts of interest}
There are no conflicts to declare.

\section*{Data availability}
The analysis code is publicly available on GitLab as the Python package \textit{patchynet} (\url{https://gitlab.com/SusanneWag/patchynet}). The raw simulation data output is available from the corresponding author upon request. All simulation parameters, model specifications, and methodological details necessary for independent verification are provided in the manuscript, enabling reproducibility through alternative simulation software.

\acknowledgments{
We gratefully acknowledge the share of CPU time on the Austrian Scientific Cluster (ASC) under Project No. fs71007. S.W. acknowledges financial support from the TU Wien funded doctorate college Biointerface. C.K. and S.W. acknowledge financial support from the Austrian Science Fund (FWF) under Project RIC 5425824.
}

\section*{Electronic Supporting Information Available}
The Supporting information details the order parameter's calculation, the method to identify the plateauing time/fraction of trajectories, supplementary analysis data, and sizes of the largest bonded cluster. 

\appendix
\section*{Appendix}
\subsection*{Basic Graph Concepts}
A \textbf{simple, undirected graph} $G$ is defined as an ordered pair $(N,E)$, where $N$ is a set of nodes, and $E$ is a set of edges. An edge is defined as a pair of distinct nodes $(n,m)$, with $n,m\in N$.\medskip

An \textbf{induced subgraph} $G'\subset G=(N,E)$ is defined by a subset of nodes $N'\subset N$ and the induced subset of edges $E'\subset E$, where $E'$ includes all edges for which both endpoints are included in $N'$.\medskip

The \textbf{degree} $\delta(i)$ of a node with index $i$ is defined as the number of incident edges to that node.\medskip

The local \textbf{clustering coefficient} $C(i)$ of a node with index $i$ is defined as the fraction of the number of edges among the next neighbors of node $i$, $e(i)$, over the maximum number of such edges \cite{Watts1998}:
\begin{equation}
    C(i)=\frac{2e(i)}{\delta(i)(\delta(i)-1)}
\end{equation}
It describes the likelihood that the bonded neighbors of node $i$ are also bonded. The local clustering coefficient is at maximum $C(i)=1$, when for each pair of neighbors of node $i$, there is an edge connecting these two neighbors. When such an edge is missing, $C(i)<1$.\medskip

The \textbf{embedding on a surface} is the drawing of a graph on that surface, such that edges intersect only at their common endpoints. An embedding of the graph in $\mathbb{R}^2$ is called a planar embedding. The graph for which such an embedding exists is called a planar graph. An embedding of the graph on the two-dimensional torus $\mathbb{T}^2$ is called a toroidal embedding, and a graph that is not planar, but for which an embedding on the torus exists, is called a toroidal graph.\medskip

The \textbf{rotation system} uniquely defines an embedding of the graph into the plane $\mathbb{R}^2$ or on the torus $\mathbb{T}^2$. It is determined by defining for each node the cyclic ordering of the incident edges connecting the node to its neighboring nodes.\medskip

\subsection*{Bonding motif detection}
We detect the bonding motifs by using the subgraph of the particle-patch graph that is induced when reducing the set of all nodes to the set including only patch nodes. In such an induced subgraph, particle nodes are removed, while patch nodes and the edges connecting these patch nodes persist. The distinct connected components of this subgraph, appearing as clusters of patch nodes, represent all bonding motifs within the particle network assembly.

\subsection*{Cyclic and Acyclic subgraphs}
We define the cyclic subgraph as the maximal induced subgraph in which all nodes are of degree $\delta(i)\ge2$ and in which all nodes have a clustering coefficient $C(i)<1$. This definition is based on the definition of the 2-core, the maximal induced subgraph in which all nodes are of degree $\delta(i)\ge2$ \cite{Seidman1983}.

To generate the cyclic subgraph, we use an \textbf{iterative pruning scheme}: In this scheme we test for each node whether both conditions, $\delta(i)\ge2$ and $C(i)<1$, hold. If a node fulfills both conditions, we keep it; when one or both conditions are broken, we remove the node from the graph. Taking the subset of nodes $N'$ for which both conditions hold, we find the induced subgraph $G'$. In the next iteration, we take the induced subgraph $G'$ of the previous iteration, until the pruning scheme converges to the cyclic subgraph, in which all nodes fulfill the conditions $\delta(i)\ge2$ and $C(i)<1$.

While the cyclic subgraph is generated using the iterative pruning scheme, the \textbf{acyclic subgraph} is defined as the complement to the cyclic subgraph with respect to the particle-patch graph, i.e. the induced subgraph taking the set of nodes that were removed in the pruning scheme.

\subsection*{Loop detection}
The loop detection is based on the embedding of the cyclic subgraph, defined by the rotation system reflecting the bond angles between the particles. We note that for particle networks that have grown beyond the periodic boundaries, the cyclic subgraph can only be embedded on a torus, while for particle networks staying within the boundaries of the simulation box, a planar embedding exists.

In a planar graph, the set of all face cycles $\mathcal{F}$ comprises all inner faces as well as the outer face $f_{out}$, which for graphs representing real particle networks can in almost all cases be identified as the face largest in size, $f_{out}=\max({\mathcal{F}})$. Looking at Fig.~\ref{fig:subgraphs} in the main text, we see that some faces are enclosed by triangle bonding motifs. Removing these faces, we find the complete collection of loops for a particle network defined as $\mathcal{L}=\{f\in\mathcal{F}: |f|>3 \}\setminus \{f_{\text{out}}\}$ for planar graphs, and for toroidal graphs as $\mathcal{L}=\{f\in\mathcal{F}: |f|>3\}$.

\subsection*{Loop size and bonding motif concentrations}
The concentration of loops of size $\ell$ is defined as the fraction of the number of loops of that size $M_{\ell}$ over all loops $M$ in the system:
\begin{align}
    c_{\ell}=\frac{M_{\ell}}{M} \text{ .}
\end{align}
Similarly the concentration of a bonding motif of type $x$ is defined as the fraction of the number of such bonding motifs $B_x$ over the number of all bonding motifs $B$ in the system:
\begin{align}
    c_{x}=\frac{B_{x}}{B} \text{ .}
\end{align}
The total number of particles $N$ can be restored from the bond motif concentrations by taking the number of patches in each bonding motif $|x|$ into account:
\begin{align}
    2N = \sum_{x}B_x|x|=B\sum_xc_x|x| \text{ ,}
\end{align}
considering that each particle has two patches. Note that such a direct relation of the number of particles and the concentration of loops does not exist, because a particle may be part of no, one or two loops, while a patch can be part of only one bonding motif.

Not only for this reason we are also interested in the fraction of particles in loops $N_L$ over all particles $N$:
\begin{align}
    n_L=\frac{N_L}{N} \text{ .}
\end{align}

\bibliography{bibtex/main.bib}

\end{document}


\title{Supporting Information - Networks of triblock ellipses: from kagome order to disordered porous networks}

\author{S.~Wagner$^1$, G.~Kahl$^1$ and C.~Karner$^1$}
\email[Correspondence email address: ]{carina.karner@tuwien.ac.at}
\affiliation{$^1$Institute for Theoretical Physics, TU Wien, Wiedner Hauptstra\ss e 8-10/136, A-1040 Wien, Austria}

\date{\today}

\maketitle

\section{Number of loops in the perfect kagome lattice}
\begin{figure}[h]
    \centering
    \includegraphics[width=0.5\linewidth]{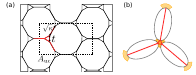}
    \caption{(a) The number of loops in the unit cell $A_{\rm uc}$ of the lattice is approximated by geometric consideration of the patchy ellipses configuration: Ellipses with major axis $2a=\sqrt{\kappa}$ arrange into six-membered loops. Small equilateral triangles with side length $t$ are inscribed by connecting the poles of the ellipses (see panel (b)).}
    \label{fig:kagome}
\end{figure}
We assess the number of loops $M^{\text{reg-}6}$ in the kagome lattice, i.e. the regular loop network with loops of size six. In order to find a suitable approximation for the related loop number density $\eta^{\text{reg-}6}$, i.e. the number of loops within the area of the unit-cell $A_{\mathrm{uc}}$, we have to consider the geometry of the lattice, as illustrated in Fig.~\ref{fig:kagome}(a). The loops are formed by ellipses with major axes of length $\sqrt{\kappa}$. As illustrated in Fig.~\ref{fig:kagome}(b), the major axes of three ellipses within the lattice do not join in a single point; instead, the poles of the ellipses are pushed apart due to the hard core repulsion and available patch bonding volume. The vertices of the three ellipses form a triangle, that we approximate as an equilateral triangle with side lengths $t$. For the unit cell area -- as indicated by the dashed rectangle in Fig.~\ref{fig:kagome}(a) -- depending on the ellipses aspect ratio $\kappa$, as well as the side lengths $t$ of the inscribed equilateral triangles, we derive:
\begin{align}
    A_{\rm uc}(\kappa,t)&=(3\sqrt{\kappa}+2\sqrt{3}t)(\sqrt{3\kappa}+2t)\\
    &=3\sqrt{3}\kappa+12t\sqrt{\kappa}+4\sqrt{3}t^2
\end{align}
As this unit cell hosts two loops, the related loop number density, $\eta^{\text{reg-}6}=\frac{2}{A_{\rm uc}}$. With parameters $\kappa=2$ and $t=0.05\sigma$, we derive this quantity to be approximately $\eta^{\text{reg-}6}\approx0.1777$. With this, the normalization constant of the global order parameter (Equ.~(4) in the main text) is expressed as $M^{\text{reg-}6}=\eta^{\text{reg-}6}A_{\rm sim-box}$, with the area of the simulation box $A_{\rm sim-box}$.

\section{Analysis of simulation trajectories}\label{sub:plateau-SI}
To distinguish simulation runs that reach a plateau from trajectories that exhibit a slow monotonic drift (creeping), we analyze trajectories of the total energy over the simulation time $E_{\mathrm{tot}}(t_i)$, with index $i=1,...,n_{\text{conf}}$ indicating a configuration. The total run time of the trajectory is $t_{\mathrm{run}}=n_{\mathrm{conf}}\cdot\tau_{\mathrm{mc}}$, where $\tau_{\mathrm{mc}}=2\times10^3$ mc-cycles. Additionally, we determine the plateauing time $t_0$, that marks the starting time for taking the ensemble-averages. The analysis consists of two steps, which rely on a small set of numerical parameters: the size of the smoothing window, the slope tolerances $k'_{\text{tol}}$ and $k''_{\text{tol}}$, and the threshold $R^2_{\mathrm{tol}}$ of the coefficient of determination. All of them were chosen empirically: we tested different parameter sets and validated the resulting classifications and plateauing times by visual inspection of the trajectories. Moderate variations of the parameters do not significantly change the overall outcome.

\subsection*{Step 1: Determining the onset of plateauing}
We determine the onset of plateauing at time $t'_{0}$ by means of linear regression on noise-filtered data over a sequence of overlapping time windows. The trajectory is smoothed by a moving average with a window size between $\Delta t_{\rm filter}=50\tau_{\mathrm{mc}}$ and $500\tau_{\mathrm{mc}}$, depending on the total length of the trajectory and chosen such that $4\cdot\Delta t_{\rm filter}<t_{\mathrm{run}}$ is fulfilled. We note that the moving average acts as a low-pass filter, suppressing high-frequency noise in the data. However, due to the truncated averaging window, the smoothing is less effective at the beginning and end of the trajectory.

We then apply a sequence of time windows $[t_i-\Delta t,t_i]$ with 90\% overlap between consecutive windows, and $\Delta t=0.5\cdot \Delta t_{\rm filter}$. Starting at $i=n_{\text{conf}}$ and proceeding backwards in time, we determine $t'_{0}$ as the last step for which the linear regression analysis on the smoothed trajectory meets the slope tolerance criterion $|k|<k'_{\mathrm{tol}}$, with $k'_{\mathrm{tol}}=0.016\, k_{\mathrm{B}}T/\tau_{\mathrm{mc}}$. A trajectory is accepted as plateauing in Step 1 if $t'_0<0.98\cdot t_{\mathrm{run}}$, and rejected if $t'_0\geq 0.98\cdot t_{\mathrm{run}}$.

Step 1 has two limitations: First, because the noise filter is imperfect, particularly at the end of the trajectory where the smoothing is less effective, residual noise may trigger the slope criterion prematurely. As a consequence, Step 1 sometimes rejects trajectories that do in fact plateau. Second, since the slope criterion is evaluated over short time windows, the tolerance $k'_{\mathrm{tol}}$ must be chosen generously to avoid picking up low-frequency noise that survives the filter. Slow, monotonic drifts (creeping behavior) that extend over the full trajectory while staying consistently below $k'_{\mathrm{tol}}$ within each time window therefore pass Step 1 undetected. For these reasons, we add a second step that (i) tests trajectories accepted in Step 1 for creeping behavior over $[t'_0,t_{\mathrm{run}}]$, and (ii) re-examines trajectories rejected in Step 1 over $[0.8\cdot t_{\mathrm{run}},t_{\mathrm{run}}]$.

\subsection*{Step 2: Testing for creeping behavior}

To test whether a trajectory exhibits creeping behavior, we apply a linear regression to the unfiltered trajectory data over a tail interval of the trajectory. In contrast to Step 1, this interval extends over a larger time period, which allows us to detect overall monotonic trends that go undetected by screening small time windows. We classify the trajectory as creeping if both conditions are met: (i) the slope exceeds a stricter tolerance, $|k|>k''_{\text{tol}}$, with $k''_{\text{tol}}=0.01\, k_{\mathrm{B}}T/\tau_{\mathrm{mc}}<k'_{\mathrm{tol}}$, and (ii) the coefficient of determination satisfies $R^2 \geq R^2_{\mathrm{tol}}$, with $R^2_{\mathrm{tol}}=0.1$. Since the regression is performed on unfiltered data, the variance of the signal is dominated by fluctuations even in the presence of a drift, and $R^2$ remains small. By setting the tolerance $R^2_{\mathrm{tol}}$ low, we accept noise in the signal, while still requiring the linear trend to be statistically noticeable, i.e., to account for at least 10\% of the variance.

The tail interval depends on the outcome of Step 1, and we distinguish two cases:
\begin{itemize}
  \item Trajectories accepted in Step 1 ($t'_0<0.98\cdot t_{\mathrm{run}}$) are tested on $[t'_0,t_{\mathrm{run}}]$. If creeping is detected, the trajectory is reclassified as not plateauing. Otherwise, it is confirmed as plateauing with $t''_0=t'_0$.
  \item Trajectories rejected in Step 1 ($t'_0\geq 0.98\cdot t_{\mathrm{run}}$) are tested on $[0.8\cdot t_{\mathrm{run}},t_{\mathrm{run}}]$. If no creeping is detected, the trajectory is reclassified as plateauing with $t''_0=0.8\cdot t_{\mathrm{run}}$. Otherwise, it remains classified as not plateauing.
\end{itemize}
For all trajectories classified as plateauing, we record the plateauing time as $t_0=t''_0$.

\section{Simulation Results - Overview}
\begin{figure}[h]
    \centering
    \begin{subfigure}{0.48\linewidth}
        \centering
        \includegraphics[width=\linewidth]{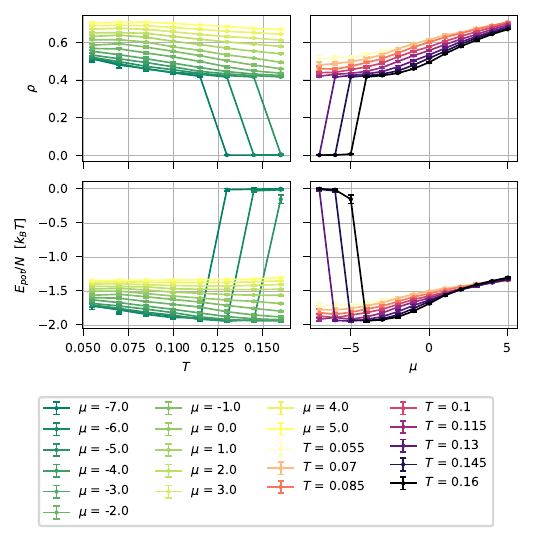}
        \caption{Density $\rho$, and energy per particle $E_{pot}/N$.}
        \label{fig:densityenergygrid}
    \end{subfigure}
    \hfill
    \begin{subfigure}{0.48\linewidth}
        \centering
        \includegraphics[width=\linewidth]{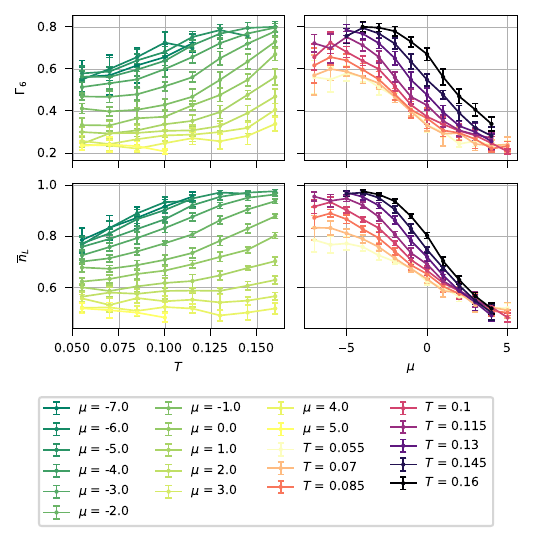}
        \caption{Global loop order parameter $\Gamma_6$ and loop particle fraction $n_L$.}
        \label{fig:loopopgrid}
    \end{subfigure}
    \caption{(a) Density $\rho$, energy per particle $E_{pot}/N$ and (b) global loop order parameter $\Gamma_6$ and loop particle fraction $n_L$ for all simulated chemical potentials and temperatures $(\mu,T)$ at which we observe condensed phases. Averages are taken over all simulation trajectories for which a plateau is reached, and error bars determined from the standard deviation across the independent trajectories.}
    \label{fig:overview}
\end{figure}

We give an overview of all simulated state points, showing the density and energy per particle in Fig.~\ref{fig:densityenergygrid}, as well as the global loop order parameter and loop particle fraction in Fig.~\ref{fig:loopopgrid}. Averages include only configurations that have reached a plateau according to our criterion, as put forward in Sec.~\ref{sub:plateau-SI}.

\section{Cross-sections at constant $\mu$ and constant $T$}
\begin{figure}[h]
    \centering
    \begin{subfigure}[b]{0.195\textwidth}
        \centering
        \includegraphics[width=\textwidth]{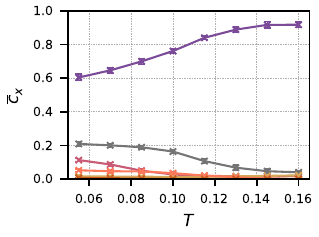}
        \label{fig:bondmotif_mu-4}
    \end{subfigure}
    \hfill
    \begin{subfigure}[b]{0.195\textwidth}
        \centering
        \includegraphics[width=\textwidth]{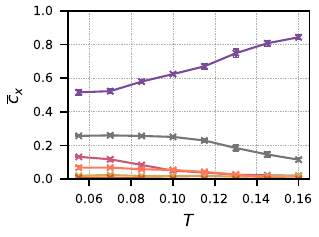}
        \label{fig:bondmotif_mu-2}
    \end{subfigure}
    \hfill
    \begin{subfigure}[b]{0.195\textwidth}
        \centering
        \includegraphics[width=\textwidth]{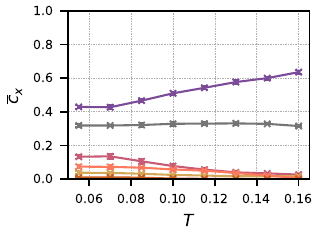}
        \label{fig:bondmotif_mu0}
    \end{subfigure}
    \hfill
    \begin{subfigure}[b]{0.195\textwidth}
        \centering
        \includegraphics[width=\textwidth]{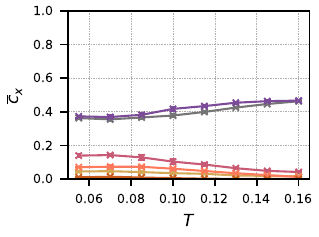}
        \label{fig:bondmotif_mu2}
    \end{subfigure}
    \hfill
    \begin{subfigure}[b]{0.195\textwidth}
        \centering
        \includegraphics[width=\textwidth]{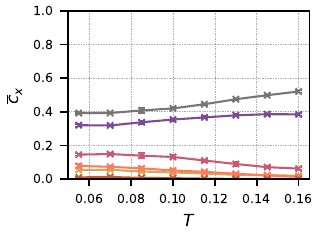}
        \label{fig:bondmotif_mu4}
    \end{subfigure}
    \\
    \begin{subfigure}[b]{0.195\textwidth}
        \centering
        \includegraphics[width=\textwidth]{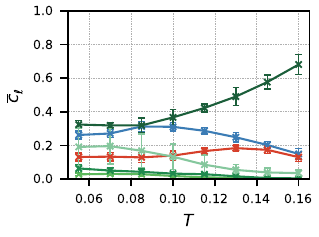}
        \caption{at $\mu=-4.0$}
        \label{fig:loopsize_mu-4}
    \end{subfigure}
    \hfill
    \begin{subfigure}[b]{0.195\textwidth}
        \centering
        \includegraphics[width=\textwidth]{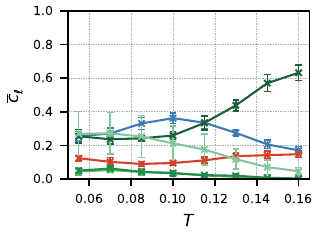}
        \caption{at $\mu=-2.0$}
        \label{fig:loopsize_mu-2}
    \end{subfigure}
    \hfill
    \begin{subfigure}[b]{0.195\textwidth}
        \centering
        \includegraphics[width=\textwidth]{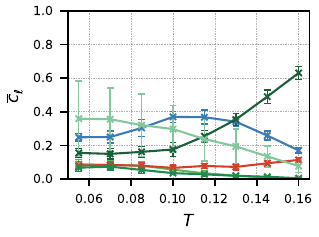}
        \caption{at $\mu=0.0$}
        \label{fig:loopsize_mu0}
    \end{subfigure}
    \hfill
    \begin{subfigure}[b]{0.195\textwidth}
        \centering
        \includegraphics[width=\textwidth]{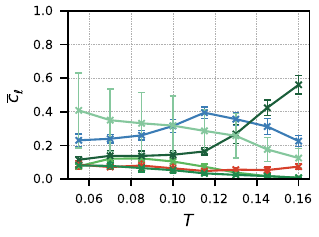}
        \caption{at $\mu=2.0$}
        \label{fig:loopsize_mu2}
    \end{subfigure}
    \hfill
    \begin{subfigure}[b]{0.195\textwidth}
        \centering
        \includegraphics[width=\textwidth]{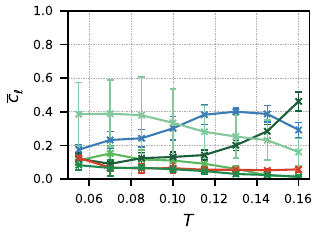}
        \caption{at $\mu=4.0$}
        \label{fig:loopsize_mu4}
    \end{subfigure}
    \caption{Cross sections of bonding motif distributions (top row) and loop size distributions (bottom row) for constant chemical potentials $\mu$ and varying temperatures. Legends for both distributions as in the main text; legend for the bond motif distributions: triangles (purple), singles (gray), pairs (pink), open triangles (yellow), N-motifs (orange), W-motifs (dark orange); legend for the loop size distributions: 4-loops (light-green), 5-loops (blue), 6-loops (dark green), 7-loops (red), 8-loops (green), 9-loops and larger (muted green).}
    \label{fig:crossmu_distributions}
\end{figure}

\begin{figure}[h]
    \centering
    \begin{subfigure}[b]{0.195\textwidth}
        \centering
        \includegraphics[width=\textwidth]{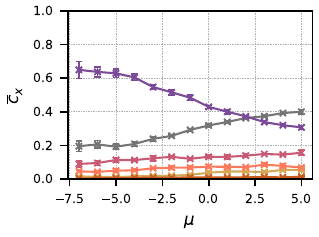}
        \label{fig:bondmotif_T0055}
    \end{subfigure}
    \hfill
    \begin{subfigure}[b]{0.195\textwidth}
        \centering
        \includegraphics[width=\textwidth]{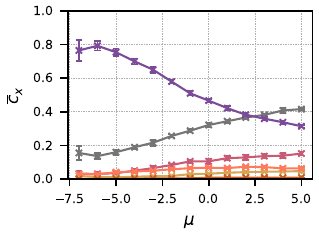}
        \label{fig:bondmotif_T0085}
    \end{subfigure}
    \hfill
    \begin{subfigure}[b]{0.195\textwidth}
        \centering
        \includegraphics[width=\textwidth]{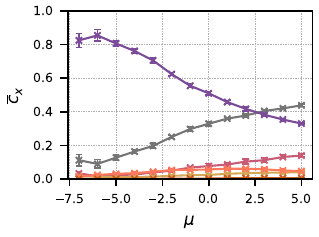}
        \label{fig:bondmotif_T01}
    \end{subfigure}
    \hfill
    \begin{subfigure}[b]{0.195\textwidth}
        \centering
        \includegraphics[width=\textwidth]{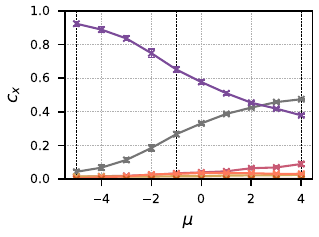}
        \label{fig:bondmotif_T013}
    \end{subfigure}
    \hfill
    \begin{subfigure}[b]{0.195\textwidth}
        \centering
        \includegraphics[width=\textwidth]{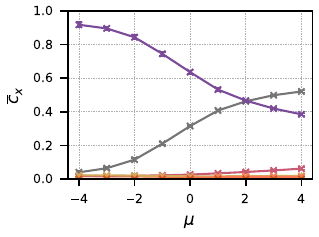}
        \label{fig:bondmotif_T016}
    \end{subfigure}
    \\
    \begin{subfigure}[b]{0.195\textwidth}
        \centering
        \includegraphics[width=\textwidth]{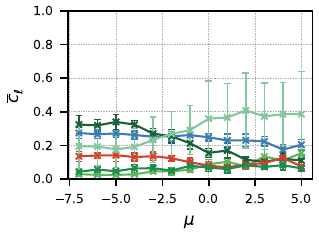}
        \caption{at $T=0.055$}
        \label{fig:loopsize_T0055}
    \end{subfigure}
    \hfill
    \begin{subfigure}[b]{0.195\textwidth}
        \centering
        \includegraphics[width=\textwidth]{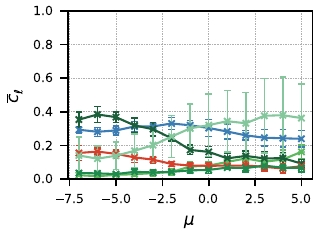}
        \caption{at $T=0.085$}
        \label{fig:loopsize_T0085}
    \end{subfigure}
    \hfill
    \begin{subfigure}[b]{0.195\textwidth}
        \centering
        \includegraphics[width=\textwidth]{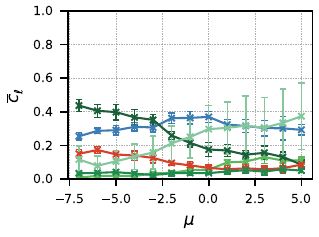}
        \caption{at $T=0.1$}
        \label{fig:loopsize_T01}
    \end{subfigure}
    \hfill
    \begin{subfigure}[b]{0.195\textwidth}
        \centering
        \includegraphics[width=\textwidth]{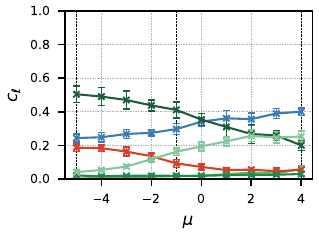}
        \caption{at $T=0.13$}
        \label{fig:loopsize_T013}
    \end{subfigure}
    \hfill
    \begin{subfigure}[b]{0.195\textwidth}
        \centering
        \includegraphics[width=\textwidth]{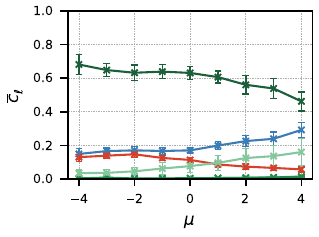}
        \caption{at $T=0.16$}
        \label{fig:loopsize_T016}
    \end{subfigure}
    \caption{Cross sections of bonding motif distributions (top row) and loop size distributions (bottom row) for constant temperatures $T$ and varying chemical potentials. Legends for both distributions as in the main text; legend for the bond motif distributions: triangles (purple), singles (gray), pairs (pink), open triangles (yellow), N-motifs (orange), W-motifs (dark orange); legend for the loop size distributions: 4-loops (light-green), 5-loops (blue), 6-loops (dark green), 7-loops (red), 8-loops (green), 9-loops and larger (muted green).}
    \label{fig:crossT_distributions}
\end{figure}

Following our discussion of two representative cross sections, one at constant $T=0.145$ and another at constant $\mu=1.0$ in the main text, we support our conclusions by providing additional cross sections for constant chemical potentials $\mu$ in Fig.~\ref{fig:crossmu_distributions} for the bonding motif and loop size distributions; as well as cross sections for constant temperature $T$ in Fig.~\ref{fig:crossT_distributions}.

\section{Nucleation}
\begin{figure}[h]
    \centering
    \includegraphics{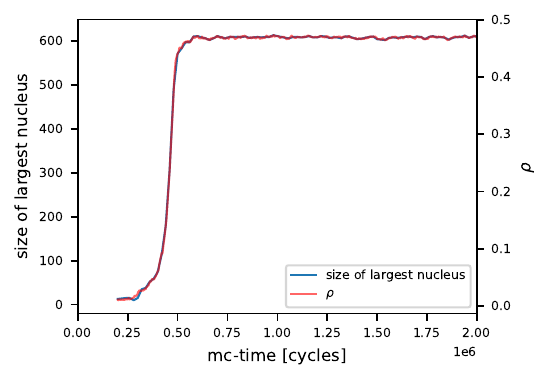}
    \caption{Size of the largest nucleus (blue) and density (red) as functions of mc-time at the state point $T=0.07$ and $\mu=-7.0$ shown for a randomly selected trajectory.}
    \label{fig:nucleus_size}
\end{figure}

At negative chemical potentials in the ranges between $\mu=-7$ to $\mu=-5$ we observe nucleation events during production run. These state points are framed in blue in Figure~4 of the main text. For these state points we observe nucleation across all twelve trajectories. 
In Figure~7 of the main text we show the density, the bonding-motif counts as well as number of different loop types as functions of mc-time. All of these quantities supply evidence that patchy ellipses do nucleate from vapor for all three temperatures. 
Here, in Figure~\ref{fig:nucleus_size} we plot the size of the largest cluster of bonded particles as a function of mc-time for the chemical potential $\mu=-7$ and the temperature $T=0.07$, alongside the density $\rho$ for a single trajectory. In this plot, we observe that the $\rho$ and the size of the largest cluster show an almost identical growth behavior, supplying evidence that nucleation proceeds by growth of the largest cluster in the system. 
